\documentclass[comsoc,10pt]{IEEEtran}
\usepackage[T1]{fontenc} 
\usepackage[utf8]{inputenc}
\usepackage{amsmath}
\usepackage[cmintegrals]{newtxmath}
\usepackage{bm} 
\usepackage{siunitx}
\usepackage{tikz}
\usepackage{pgfplots}
\usepackage{subcaption}
\pgfplotsset{compat=1.6}
\usetikzlibrary{external}
\usetikzlibrary{shapes, arrows,graphs}
\tikzset{external/force remake}
\usepackage{pgfplots}
\pgfplotsset{compat=1.12}
\usepgfplotslibrary{external,polar}
\usepackage{soul}
\usepackage[final,xcolor={red,dvpipdf}]{changes}

\definechangesauthor[name={Nils Johannsen}, color=blue]{NJ}
\definechangesauthor[name={Nikolai Peitzmeier}, color=red]{NP}

\colorlet{Changes@Color}{red}

\usepackage{mathrsfs}
\graphicspath{{./pic/}}
\def\BibTeX{{\rm B\kern-.05em{\sc i\kern-.025em b}\kern-.08em
		T\kern-.1667em\lower.7ex\hbox{E}\kern-.125emX}}

\IEEEoverridecommandlockouts

	\title{Single-Element Beamforming using Multi-Mode Antenna Patterns}
\author{
	\IEEEauthorblockN{Nils L. Johannsen and Peter A. Hoeher, \IEEEmembership{Fellow, IEEE}}\\
	\IEEEauthorblockA{\textit{Chair of Information and Coding Theory},
		\textit{Kiel University},
		Kiel, Germany, \{nj,ph\}@tf.uni-kiel.de}\\
	}

\usepackage{environ}
\makeatletter
\newsavebox{\measure@tikzpicture}
\NewEnviron{scaletikzpicturetowidth}[1]{%
	\def\tikz@width{#1}%
	\begin{lrbox}{\measure@tikzpicture}%
		\BODY
	\end{lrbox}%
	\pgfmathparse{#1/\wd\measure@tikzpicture}%
	\BODY
}
\makeatother

\begin{document}
	\pgfplotsset{compat=1.3,
		every axis legend/.style={
			y tick label style={/pgf/number format/1000 sep=},					
			x tick label style={/pgf/number format/1000 sep=},
			z tick label style={/pgf/number format/1000 sep=}
	}}
	
	\maketitle
	\begin{abstract}
		Multi-mode antennas have recently been studied for communication as
		well as localization purposes.
		In this work, the capabilities provided by a single planar multi-mode
		radiator as a steerable multi-port antenna are explored.
		As an original contribution, the radiation characteristics of individual
		groups of modes of the single radiator are combined to optimize
		beamforming performance.
		\added{Three possible codebook realizations resulting
		from the optimization of gain, element factor,
		and, as a new criterion, gain by element factor are studied and compared.} 
		\added{Numerical results show that the gain can be enhanced by about 3~dB at some angles.} 
	\end{abstract}
	
	\section{Introduction}
	\label{sec:Introduction}

	In the area of antenna design, the excitation of a given or designed aperture with respect to a low reflection coefficient and high efficiency is a major challenge \cite{ZWTYR19,GMWZP16}. 
	This holds especially in the field of multi-mode antennas \cite{ChWa15, LMAH16}, also called modal antennas. 
	The theory of characteristic modes does not give answers towards engineering antenna parameters like gain, side-lobe level, etc. 
	In order to optimize the effectiveness of these antennas, MIMO signal processing is an option. 
	Several planar multi-mode antenna elements can be assembled to become a massive MIMO multi-mode antenna array suitable for base stations and access points, whereas a single multi-mode antenna element is suitable for compact mobile equipment \cite{MaMa16}.
	Hence, MIMO signal processing is either applicable to the entire array, or to a single element.
	
	In the numerical results presented below, focus will be on the planar multi-mode antenna design presented in \cite{MaMa16,PeMa18}. So far, this antenna type has been investigated for communication and localization purposes.
	In \cite{PeMa19a} an upper bound of the achievable number of orthogonal ports is discussed.
	 In \cite{HoDo17,HMDP17} an ultra-wideband (UWB) communication system targeting data rates of 100 Gbps and beyond at frequencies below 10~GHz is proposed.
	In this context, the different radiation patterns of the antenna ports are assumed to separate different radiation directions of an antenna array, and therefore improve the capability of multi-stream data processing. Emphasis is on increasing the overall sum-rate of the system. Contrary, in \cite{APDHD18,PAZJDH19} multi-mode antennas are used for angle-of-arrival estimation for the purpose of precise positioning. In the latter contributions, a single multi-mode antenna element is modeled as a uniform linear array (ULA). Besides communication and localization applications, multi-mode antennas are believed to be applicable in the context of radar applications as well.
	
	\added{This work focuses on the ability of using orthogonal multi-mode antenna patterns to allow beamforming using a single radiator element.}
	Different patterns corresponding to the ports of a multi-mode antenna are combined in order to achieve an optimization of the overall radiation characteristics by making use of linear precoding independently of the antenna design process.
	An optimization in terms of achievable \emph{gain} is introduced. To reduce the impact of potential interferers, the \emph{element factor} is used as a design parameter.
	To merge \emph{gain} and \emph{element factor}, a new design parameter, \emph{gain by element factor}, is introduced.
	The number of scheduled ports is investigated per optimization criterion.
	\added{Since digital beamforming increases cost and power consumption, different codebooks for different precoding setups including hybrid and analog precoding are studied and their performance is compared.}
	\emph{Digital mode combining}, \added{as a digital beamforming strategy,} provides maximum flexibility and hence is proposed as a benchmark. 
	In hardware-limited scenarios, \emph{hybrid} and purely \emph{analog mode combining} offer a trade-off between complexity and efficiency. 
	\emph{Mode selection}, \added{as a beamforming strategy which only selects one mode,} is the simplest alternative, at the expense of gain performance.
	
	\added{This investigation is discussed for the radiation at center frequency. However, when assuming a multi-carrier transmission system like OFDM and a digital beamforming architecture, similar results can be expected for the full bandwidth, since orthogonality of radiation is provided by the multi-mode antenna designed in \cite{MaMa16}.}
	
	\section{Basics of Weighted Mode Combining}
	\label{sec: SystemModel}
\begin{figure}[h!]
		\centering
		\includegraphics[width=0.75\columnwidth]{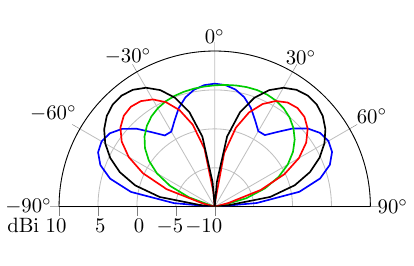}
		\caption{Radiation patterns in y-z-plane of the 4-port prototype antenna from \cite{MaMa16}.}
		\label{fig: Radiation4Modes}
	\end{figure}

	Let us consider a multi-mode antenna design with orthogonal groups of surface current distributions according to the theory of characteristic modes.
	Regarding UWB and multi-mode capability, an electrically large aperture is necessary to enable different current distributions at the same time, as proposed in \cite{MaMa16}.
	As the feeding network connects feeding points and antenna input ports in a sophisticated manner \cite{MaMa16,PeMa18}, each physical port corresponds to a unique group of modes, which can be excited individually.
	Since the groups of modes are orthogonal among each other, the ports are uncoupled.
	Let the electric field components be denoted as $F_\phi$ and $F_\theta$, respectively, where $\phi$ is the azimuth angle and $\theta$ the elevation.
	If 
	\begin{equation}
		\frac{1}{2 Z_0}\oint_\Omega \big| F_\phi(\phi,\theta)\big|^2 + \big| F_\theta(\phi,\theta)\big|^2 d\Omega = 1
	\end{equation}
	is fulfilled and the efficiency of the antenna is set to be one, the radiated power of each group of modes corresponds to the input power provided at the associated antenna port.
	Subsequently, the gain $G_m$ of the $m$-th port and mode group $m$ in a certain direction $\phi,\theta$ can be calculated by 
	\begin{equation}
		G_m(\phi,\theta) = \frac{4\pi}{2Z_{0}}\Big[\big| F_{\phi,m}(\phi,\theta)\big|^2 + \big| F_{\theta,m}(\phi,\theta)\big|^2 \Big],
		\label{eq: EIRP}
	\end{equation}
	where $1 \le m \le M$.
	\deleted{The equivalent isotropic radiated power (EIRP) of this port $m$ is defined as \cite{DoHo18}}
	The corresponding radiation patterns at the design frequency of the individual ports in the y-z-plane are shown in Fig.~\ref{fig: Radiation4Modes} for the 4-port prototype antenna under investigation ($M = 4$).
	
	When using a single multi-mode antenna element, the \emph{gain} in azimuth $\phi$ and elevation $\theta$ can be calculated by using \eqref{eq: EIRP} and introducing complex-valued weighting coefficients $c_m$ as
	\begin{equation}
		G(\phi,\theta)=\frac{4\pi}{2Z_{0}}\bigg[\Big|\sum_{m} c_mF_{\phi,m}(\phi,\theta)\Big|^2 + \Big|\sum_{m} c_m F_{\phi,m}(\phi,\theta)\Big|^2\bigg],
		\label{eq: subarrayGain}
	\end{equation}
	subject to the constraint $\sum_{m}|c_m|^2 = 1$.
	Note that opposed to classical array processing, the achievable gain of single-element beamforming cannot be estimated as a function of the number of modes.
	In order to optimize the radiation in terms of \emph{gain} (or other criteria) at a given angle, the complex weighting coefficients $c_m$ have to be designed such that the electric-field components representing the different modes interfere constructively. Since maximizing the \emph{gain} of an antenna neglects the impact of potential interferers located at other angles, the \emph{element factor} of the antenna is of interest. The \emph{element factor} $E\!F$ of any antenna can be described by \added{dividing the gain pattern given in} \eqref{eq: EIRP} by its maximum:
	\begin{equation}
		E\!F(\phi,\theta) = \frac{G(\phi,\theta)}{\max_{\phi,\theta}\{\textrm{G}(\phi,\theta)\}}.
		\label{eq: ElementFactor}
	\end{equation}
	The corresponding values only represent the portion of power radiated to the desired direction.
	Consequently, information about the portion of power in the direction of interest gets lost and is not included in the criterion itself.
	Therefore, as a new criterion, the \emph{gain by element factor} is introduced as
	\begin{equation}
		GE\!F (\phi,\theta) = G(\phi,\theta)\cdot E\!F(\phi,\theta).
		\label{eq: GainByElementFactor}
	\end{equation}
	By combining \emph{gain} and \emph{element factor} this criterion includes distortions by interferers as well as power maximization. 
	
	\section{Codebook Optimization}
	\label{sec: UserScheduling}
	Towards the optimization of the gain (or other criteria) of a single multi-mode antenna element, the complex weighting coefficients $c_m$ and electric field components $F_\phi$ and $F_\theta$ are rewritten in matrix notation as
	\begin{equation}
		\bm{F} 
		= \begin{bmatrix}
				F_{\phi,1} & \dots & F_{\phi,M} \\
				F_{\theta,1} & \dots & F_{\theta,M} \\
			\end{bmatrix},\quad \bm{c} = \begin{bmatrix}
			c_1&\dots&c_M\\
		\end{bmatrix}^T.
	\label{eq: PatternMatrixDefinition}
\end{equation}
	Note that for better readability the angular arguments $\phi$ and $\theta$ are neglected.
	\deleted{Then,} The problem structure \added{(gain optimization)} is defined by maximizing \eqref{eq: subarrayGain} using the matrix notation from \eqref{eq: PatternMatrixDefinition} as 
	\begin{equation}
		\arg \max_{c_{m}} \bigl\{ G(\phi,\theta) \bigr\} = \arg\max_{\bm{c}}\bigl\{|\bm{F}\cdot \bm{c}|^2\bigr\}\quad \textrm{s.t.}\quad |\bm{c}|^2 \leq 1.
		\label{eq: Objective}
	\end{equation}
	\added{This convex optimization problem can be solved by standard techniques. Similarly, \eqref{eq: ElementFactor} and \eqref{eq: GainByElementFactor} can be formulated as an optimization problem and solved accordingly.}
	\deleted{Since the optimization is performed by applying an exhaustive search iterating through a limited amplitude and phase range, a best value is always reached.}
	Upon designing codebooks, several approaches are feasible. In the case of digital beamforming, each port of the multi-mode antenna is connected to the digital processing unit via a full radio frequency (RF) chain and digital-to-analog conversion. This is referred to as \emph{digital mode combining} in the remainder. 
	Digital mode combining provides best performance due to maximum flexibility for the weighting coefficients. According to \eqref{eq: Objective} and given the constraint $\sum_{m}|c_m|^2 = 1$, the coefficient vector $\mathbf{c}$ can be written as
	\begin{equation}
		\mathbf{c} = \frac{1}{\sqrt{C}}\big[a_1\,e^{j\alpha_1},\dots, a_m\, e^{j\alpha_m},\dots,a_M \, e^{j\alpha_M}\big]^T,
	\end{equation} 
	with $C = {\sum\limits_{m=1}^{M}a_i},\quad a_i \epsilon\Re^{\ge0}$ being a normalization constant to fit the power constraint.
	When trying to reduce the hardware effort, hybrid beamforming structures come into play. Different approaches are possible. Starting with a simple structure, only the port providing the best performance in a certain direction can be selected:
	\begin{equation}
	\mathbf{c} = \big[0,\dots, 0,1,0,\dots 0\big]^T.
	\end{equation}
	 This strategy has been suggested in \cite{HoDo17} and is called \emph{mode selection} subsequently.
	
	Alternatively, adjustable phase shifters can be used to connect the RF chain with all ports of the antenna element. In this case, all ports are active by definition. The optimized phases can be stored in a codebook, according to
	\begin{equation}
	\mathbf{c} = \frac{1}{\sqrt{M}} \big[e^{j\alpha_1},\dots,  e^{j\alpha_m},\dots, e^{j\alpha_M}\big]^T.
	\end{equation} 
	 Due to pure analog beamforming, this version is called \emph{analog mode combining}.
	
	A more flexible version is to implement the phases by using fixed phase shifters.
	Triggered by a processor, selected ports and the desired phases can be chosen. Because of the hybrid nature, we call this technique \emph{hybrid mode combining}. Consequently,
	\begin{equation}
	\mathbf{c} = \frac{1}{\sqrt{N_{\textrm{C}}}}\big[a_1\,e^{j\alpha_1},\dots, a_m\, e^{j\alpha_m},\dots,a_M \, e^{j\alpha_M}\big]^T,
	\end{equation}
	where before normalization the amplitudes $a_m$ are either zero or one.
	$N_\textrm{C}$ represents the number of selected ports in the current setup.
	Since an arbitrary number of ports are selected in order to optimize the overall \emph{gain} (or other criteria), this case is more flexible and hence an improved performance can be obtained.
	
	\section{Numerical Results}
	\label{sec: NumericalResults}
		\begin{figure}[h]
			\centering
			\includegraphics[width=0.85\columnwidth]{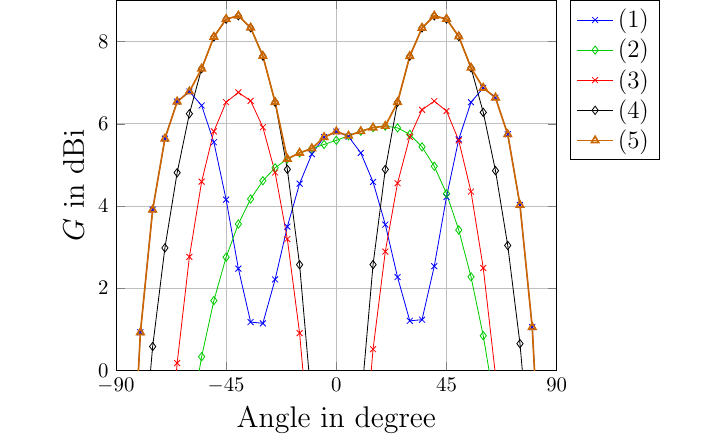}
			\caption{Achievable gain by different mode combining techniques. (1) Digital mode combining, (2) hybrid mode combining, (3) analog mode combining, (4) mode selection.}
			\label{fig: DrawReferenceEIRP}
		\end{figure}
		
	As a benchmark, let us start with \emph{mode selection} as defined above.
	In Fig.~\ref{fig: DrawReferenceEIRP}, the individual gains of the four ports are plotted in curves (1)-(4) for illustrative purposes.
	\emph{Mode selection} can be achieved by appointing the port providing the largest gain at a certain angle.
	The corresponding gain is depicted in curve (5).
	
	\begin{figure*}
		\begin{subfigure}{\textwidth}
			\centering
			\includegraphics[width=0.8\columnwidth]{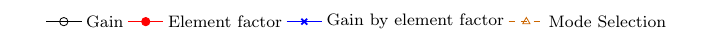}
		\end{subfigure}
		\begin{subfigure}{0.32\textwidth}
			\includegraphics[width=\columnwidth]{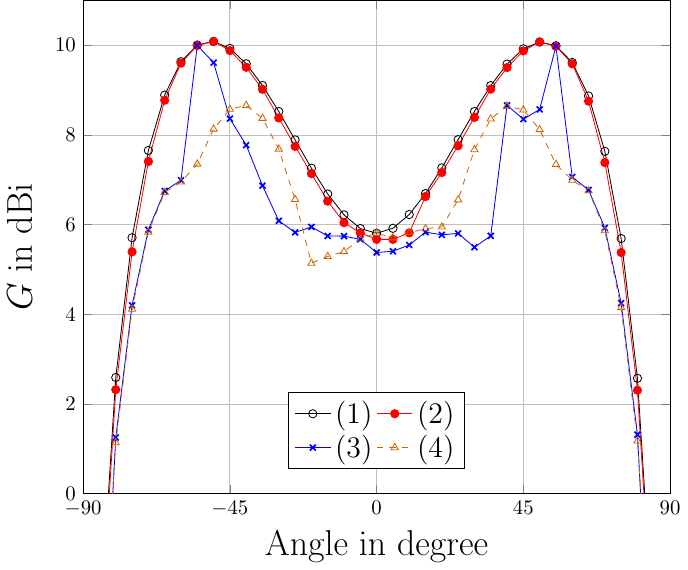}
			\caption{Achievable gain.}
			\label{fig: DrawGain3Criteria}
		\end{subfigure}
		\begin{subfigure}{0.32\textwidth}
			\includegraphics[width=\columnwidth]{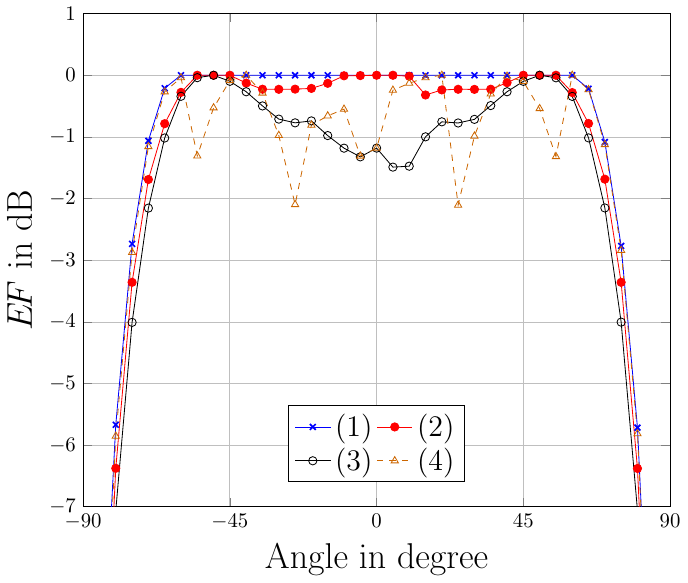}
			\caption{Achievable element factor.}
			\label{fig: DrawEF3Criteria}
		\end{subfigure}
		\begin{subfigure}{0.32\textwidth}
			\includegraphics[width=\columnwidth]{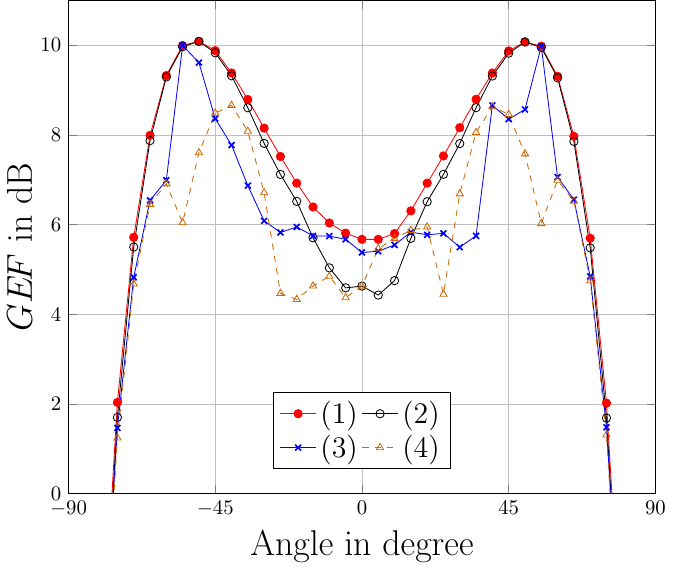}
			\caption{Achievable gain by element factor.}
			\label{fig: DrawG2EF3Criteria}
		\end{subfigure}
		\caption{Performance of the three codebooks for gain, element factor, gain by element factor, and mode selection (as reference).}
		\label{fig: 3Criteria}
	\end{figure*}
	In order to compare the performance of the different optimization criteria, the performances are plotted for each criterion.
	The achieved gain is depicted in Fig.~\ref{fig: DrawGain3Criteria}.
	In curve (1) the \emph{gain} is maximized.	
	Close to this performance comes the achieved gain of the codebook designed to fit the new criterion \emph{gain by element factor}, as depicted in curve (2).
	Finally, the gain of the codebook designed to fit the best \emph{element factor} is presented in curve (3).
	As can be seen, the achieved gain is fluctuating.
	In some points, it is even worse than the gain of the \emph{mode selection scheme}, given in curve (4).

	Fig.~\ref{fig: DrawEF3Criteria} shows that the optimal performance of the \emph{element factor}, namely  $E\!F=1$, is reached for the corresponding codebook in nearly the entire angular region.
	The element factor is presented by curve (1).
	A good performance is also reached by using the codebook designed for the \emph{gain by element factor}, which is demonstrated in curve (2).
	The codebook optimized to fulfill maximum \emph{gain} requirements is represented by curve (3).
	It has the weakest performance in terms of element factor, which results in larger interference power to neighboring users.

	When comparing the \emph{gain by element factors} of the three codebooks, as expected, the performance of codebook optimized to this criterion performs best, as is shown by curve (1).
	In the angular region below -30$^\circ$ and above 30$^\circ$, the codebook for maximum \emph{gain} achieves a similar performance, given by curve (2).
	In contrast, in the central region around 0$^\circ$, its performance is roughly 1.5~dB less compared to the achievements of the codebook in curve (1).
	Here, the performance of the codebook designed to fit the \emph{element factor}, curve (3), is better.

		\begin{figure}
			\centering
			\includegraphics[width=0.7\columnwidth]{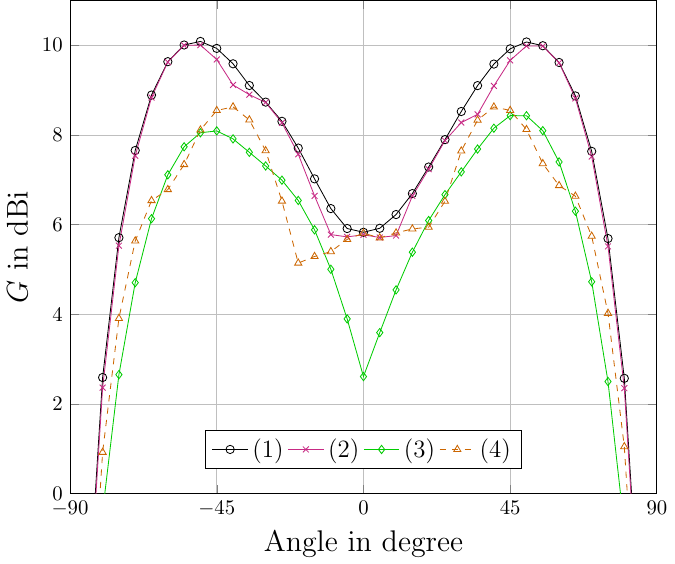}
			\caption{Achievable gain by different mode combining techniques. (1) Digital mode combining, (2) hybrid mode combining, (3) analog mode combining, (4) mode selection.}
			\label{fig: DrawEIRPOptimization}
		\end{figure}

	In Fig.~\ref{fig: DrawEIRPOptimization}, the gain achieved by the different \emph{mode combining} techniques is shown.
	\emph{Digital mode combining}, featured in curve (1), provides the best results.
	An exhaustive search considering all possible gain factors and phase angles of the weighting coefficients $\mathbf{c}$ is conducted.
	It enables up to 10\,dBi, which is an improvement of close to 3\,dB in some directions.
	Furthermore, the gain is improved nearly for the entire angular domain.
	The gain achieved by just using phase shifters and switching between contributing ports is shown in curve (2).
	Compared to digital mode combining, the loss of \emph{hybrid mode combining} is fairly small, but hardware complexity is significantly reduced.
	\emph{Analog mode combining} has the weakest gain, as shown by curve (3). 
	Curve (4) is the reference plot defined by \emph{mode selection}.	
	
	\begin{figure}[h]
		\begin{subfigure}{0.05\columnwidth}
			\includegraphics[width=\columnwidth]{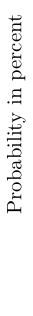}
		\end{subfigure}
		\begin{subfigure}{0.37\columnwidth}
			\includegraphics[width=\columnwidth]{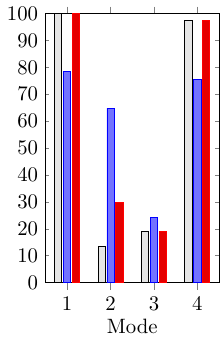}
			\caption{Incidence in percent each port is employed. }
			\label{fig: DrawBarChartModes}
		\end{subfigure}
		\begin{subfigure}{0.05\columnwidth}
			\includegraphics[width=2\columnwidth]{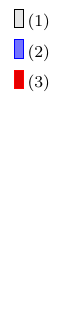}
		\end{subfigure}
		\begin{subfigure}{0.37\columnwidth}
			\includegraphics[width=\columnwidth]{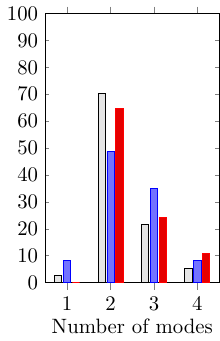}
			\caption{Number of ports employed per angle.}
			\label{fig: DrawBarChartNoOfModes}
		\end{subfigure}

			\caption{{The bar charts are generated from the digital-mode-combining codebook for 37 equidistant angles over the 180 degree hemisphere. The incidences are shown for the three codebooks designed to fit (1) gain, (2) element factor,  (3) gain by element factor.}}
			\label{fig: BarPlots}
	\end{figure}	
	In order to draw conclusions about the necessity of the individual ports and combinations, codebooks providing the best performances are investigated.
	In Fig.~\ref{fig: DrawBarChartModes} the incidence for all ports available is shown.
	As can be observed, the first port, corresponding to the gain of the blue curve in Fig.~\ref{fig: DrawReferenceEIRP}, is employed at all angles for the two of the three codebooks depending on the gain of the system. 
	This is remarkable, since it provides the largest individual gain in only very sparse cases, as compared to the fourth port shown in black.
	The radiation pattern corresponding to Port 4 is scheduled in most angles as well.
	The codebook designed to fit the \emph{element factor} employs the patterns corresponding to the Ports 1 and 4 less often. It more frequently enables Port 2 to reduce the impact of interferers, which is employed in the codebook towards \emph{gain} very sparsely. Port 3 is generally used most rarely. 	
 	Regarding the required hardware and precoding effort, the number of scheduled modes per angle segment is evaluated in Fig.~\ref{fig: DrawBarChartNoOfModes}. As can be seen, in nearly all cases two or more modes are employed in order to achieve the performance shown in Fig.~\ref{fig: DrawEIRPOptimization}, curve (1). To address the defined criteria, in roughly 70\% of the angles two modes are sufficient. Here, the codebook designed to fit the \emph{element factor} provides the greatest distribution in the usage of modes. Only in 50\% of the angles exactly two modes are employed. In about 10\% of the angles only one port is sufficient. Looking at the codebook designed to fit the \emph{gain by element factor}, always two or more modes are required. In 10\% of the angles even all four modes are selected.

	\section{Conclusions}
	\label{sec: Conclusions}
		In this work, criteria for the optimization of baseband processing given a planar multi-mode radiator have been discussed.
		A new optimization criterion, gain by element factor, has been introduced.
		Optimizing codebooks with respect to the new criterion results in a solid performance.
		As an original contribution, single-element beamforming by using the orthogonal properties of a planar multi-mode antenna has been presented.
		It is shown that the gain of the antenna element in a certain direction is improved by combining the modes.
		Additionally, the impact of interferers located at other angles can be reduced at the same time by using the new gain by element factor.
		
	\section{Acknowledgement}
		This work is supported by the German Research Foundation (DFG) within the priority program SPP 1655 under contract no.~HO 2226/14-2.
		The cooperation with Prof. Dirk Manteuffel and Nikolai Peitzmeier within this program is highly recognized. 
		Many thanks also to Bolla Vijender for initial contributions within his master thesis at Kiel University.

	\bibliographystyle{IEEEtran}
	\bibliography{IEEEabrv,../../../../bibFiles/ICTabrv.bib,../../../../bibFiles/literature.bib}

\begin{thebibliography}{10}
\providecommand{\url}[1]{#1}
\csname url@samestyle\endcsname
\providecommand{\newblock}{\relax}
\providecommand{\bibinfo}[2]{#2}
\providecommand{\BIBentrySTDinterwordspacing}{\spaceskip=0pt\relax}
\providecommand{\BIBentryALTinterwordstretchfactor}{4}
\providecommand{\BIBentryALTinterwordspacing}{\spaceskip=\fontdimen2\font plus
\BIBentryALTinterwordstretchfactor\fontdimen3\font minus
  \fontdimen4\font\relax}
\providecommand{\BIBforeignlanguage}[2]{{%
\expandafter\ifx\csname l@#1\endcsname\relax
\typeout{** WARNING: IEEEtran.bst: No hyphenation pattern has been}%
\typeout{** loaded for the language `#1'. Using the pattern for}%
\typeout{** the default language instead.}%
\else
\language=\csname l@#1\endcsname
\fi
#2}}
\providecommand{\BIBdecl}{\relax}
\BIBdecl

\bibitem{ZWTYR19}
Z.~{Zhou}, Z.~{Wei}, Z.~{Tang}, Y.~{Yin}, and J.~{Ren}, ``Compact and wideband
  differentially fed dual-polarized antenna with high common-mode
  suppression,'' \emph{{IEEE} Access}, vol.~7, pp. 108\,818--108\,826, 2019.

\bibitem{GMWZP16}
Y.~Gao, R.~Ma, Y.~Wang, Q.~Zhang, and C.~Parini, ``Stacked patch antenna with
  dual-polarization and low mutual coupling for massive {MIMO},'' \emph{{IEEE}
  Trans. Antennas Propag.}, vol.~64, no.~10, pp. 4544--4549, Oct 2016.

\bibitem{ChWa15}
Y.~Chen and C.~F. Wang, \emph{Characteristic Modes: Theory and Applications in
  Antenna Engineering}.\hskip 1em plus 0.5em minus 0.4em\relax Wiley, 2015.

\bibitem{LMAH16}
B.~K. Lau, D.~Manteuffel, H.~Ari, and S.~V. Hum, ``Guest editorial: Theory and
  applications of characteristic modes,'' \emph{{IEEE} Trans. Antennas
  Propag.}, vol.~64, no.~7, pp. 2590--2594, 2016.

\bibitem{MaMa16}
D.~Manteuffel and R.~Martens, ``Compact multimode multielement antenna for
  indoor {UWB} massive {MIMO},'' \emph{{IEEE} Trans. Antennas Propag.},
  vol.~64, no.~7, pp. 2689--2697, Jul. 2016.

\bibitem{PeMa18}
N.~Peitzmeier and D.~Manteuffel, ``Selective excitation of characteristic modes
  on an electrically large antenna for {MIMO} applications,'' in \emph{Proc.
  European Conf. Antennas Propag. (EuCAP)}, London, UK, Apr. 2018.

\bibitem{PeMa19a}
------, ``Upper bounds and design guidelines for realizing uncorrelated ports
  on multi-mode antennas based on symmetry analysis of characteristic modes,''
  \emph{{IEEE} Trans. Antennas Propag.}, pp. 3902--3914, Mar. 2019.

\bibitem{HoDo17}
P.~A. Hoeher and N.~Doose, ``A massive {MIMO} terminal concept based on
  small-size multi-mode antennas,'' \emph{Trans. Emerging Telecommunications
  Technologies}, vol.~28, no.~2, p. e2934, Feb. 2017.

\bibitem{HMDP17}
P.~A. Hoeher, D.~Manteuffel, N.~Doose, and N.~Peitzmeier, ``Ultra-wideband
  massive {MIMO} communications using multi-mode antennas,'' \emph{Frequenz},
  vol.~71, no. 9-10, pp. 429--448, Sep. 2017.

\bibitem{APDHD18}
S.~A. Almasri, R.~P\"ohlmann, N.~Doose, P.~A. Hoeher, and A.~Dammann,
  ``Modeling aspects of planar multi-mode antennas for direction-of-arrival
  estimation,'' \emph{{IEEE} Sensors J.}, vol.~19, no.~12, pp. 4585--4597, Jun.
  2019.

\bibitem{PAZJDH19}
R.~P\"ohlmann, S.~Almasri, S.~Zhang, T.~Jost, A.~Dammann, and P.~Hoeher, ``On
  the potential of multi-mode antennas for direction-of-arrival estimation,''
  \emph{{IEEE} Trans. Antennas Propag.}, vol.~67, no.~5, pp. 3374--3386, May
  2019.

\end{thebibliography}
	
\end{document}